\documentclass[doublecol]{epl2} 

\usepackage{graphicx}
\usepackage{booktabs}

\usepackage{amsmath, amssymb}
\usepackage{textcomp}
\usepackage{subfigure}
\usepackage{siunitx}
\usepackage{color}
\usepackage[final]{pdfpages}

\definecolor{linkcolor}{rgb}{0,0,0.6} 
\usepackage[pdftex,colorlinks=true, pdfstartview=FitV, linkcolor= linkcolor, citecolor= linkcolor, urlcolor= linkcolor, hyperindex=true,hyperfigures=true]{hyperref} 

\begin{document}

\title{\bf Energy flow between two hydrodynamically coupled particles kept at different effective temperatures
}
\author{A. B\'{e}rut, A. Petrosyan, S. Ciliberto}
\institute{Universit\'{e} de Lyon \\
{Laboratoire de Physique, \'{E}cole Normale Sup\'{e}rieure de Lyon, CNRS UMR5672}\\
46, All\'{e}e d'Italie, 69364 Lyon Cedex 07, France
}
\date{today}
\begin{abstract}
{We measure the energy exchanged between two hydrodynamically coupled micron-sized Brownian particles trapped in water by two optical tweezers. The system is driven out of equilibrium  by random forcing the position of one of the two particles. The forced particle behaves as it has an ``effective temperature'' higher than that of  the other bead. This driving modifies the equilibrium variances and cross-correlation functions of the bead positions: we measure an energy flow between the particles and an instantaneous cross-correlation, proportional to the effective temperature difference between the two particles. A model of the interaction which is based on classical hydrodynamic coupling tensors is proposed. The theoretical and experimental results are in excellent agreement.}
\end{abstract}
\maketitle

The energy flux between two micro-systems kept at different temperatures and coupled only by thermal fluctuations plays an important role in out of equilibrium thermodynamics. For this reason it has been widely studied theoretically \cite{Deridda,Jarz2004,VandenBroeck,Visco,Dhar2007,Gas2009,evans_temp,Hanggi2011,Villamania}, but only a few experiments \cite{ref:Ciliberto,Pekola_1} have analyzed this problem. Furthermore in all of these studies the systems where coupled by conservative forces and the  dissipative coupling have never been considered. This is however a very important case because the coupling of two close Brownian particles is dominated by their hydrodynamic interactions in low Reynold-number regimes. These  interactions, which have been widely studied in fluid at thermal equilibrium starting from hydrodynamic calculations \cite{ref:Jeffery,ref:Onishi,ref:Batchelor}, play an important role in various physical situations.
For example, the indirect interactions mediated by the solvent modify the Brownian diffusion of two particles \cite{ref:Crocker,ref:Grier}, and gives rise to an anti-correlation at finite time between the displacements of two trapped particles, which has been studied both experimentally and numerically \cite{ref:Quake,ref:Bartlett,ref:Ou-Yang,ref:HongRu}. Systems with arrays of more than two trapped particles coupled by hydrodynamic interactions show complex dynamics and can behave as an elastic medium \cite{ref:Polin,ref:Cicuta3,ref:Cicuta2,ref:Herrera}. The hydrodynamic coupling is also responsible for the synchronisation of colloidal oscillators which can be linked to collective motions of biological systems like cilia or flagella \cite{ref:Cicuta, ref:Padgett,ref:DiLeonardo,ref:Cicuta4}, and for the pair-attractions of particles driven on a circular ring \cite{ref:Roichman,ref:Kimura}. 
Despite the variety of situations, the interactions between particles trapped at different temperatures were not studied, due to the difficulty of achieving a high temperature difference on very small scales. \\
The purpose of this letter is to study how the equilibrium statistical properties are modified by the hydrodynamic coupling between two particles trapped at different effective kinetic temperatures.  The main results of our investigation concern the energy flux and the positions correlation functions. The experimental results are compared to those of an analytical model based on classical hydrodynamic coupling tensor. The ``effective temperature difference'' is produced by random forcing the positions of one of the two particles. This is a technique that have been used in two experiments on single particle,   which have shown that the random forcing can be indeed assimilated to an ``effective temperature'' \cite{ref:Petrov, ref:Solano_EPL_noise}.


In order to study these non-equilibrium interactions we use the following experimental setup: a laser beam (wavelength \SI{532}{\nano\meter}) is separated in two beams with crossed polarizations so that there is no interference between them. A custom-built vertical optical tweezers with an oil-immersion objective (HCX PL. APO $63\times$/$0.6$-$1.4$) is used to focus each beam which creates a quadratic potential well where a silica bead (radius $R = \SI{1}{\micro\meter} \pm 5\%$) is trapped. One of the beams goes through an acousto-optic deflector (AOD) that allows to switch the position of the trap very rapidly (up to \SI{1}{\mega\hertz}). The beads are dispersed in bidistilled water at low concentration to avoid interactions with multiple other beads. The solution of beads is contained in a disk-shaped cell (\SI{18}{\milli\meter} in diameter, $\SI{1}{\milli\meter}$ in depth). The beads are trapped at \SI{15}{\micro\meter} above the bottom surface of the cell. The position of the beads is tracked by a fast camera with a resolution of \SI{115}{\nano\meter} per pixel, which after treatment gives the position with an accuracy greater than \SI{5}{\nano\meter}. The trajectories of the bead are sampled at \SI{800}{\hertz}. The stiffness of the traps $k$ is proportional to the laser intensity\footnote{It can be modified by turning an half-wave plate placed before the polarization separation or by adding neutral density filters on the beams trajectory.} and is typically about \SI{4}{\pico\newton/\micro\meter}. The two particles are trapped on a line (called ``x axis'') and separated by a distance $d$ which is tunable. For all the distances used (between $2.8$ and \SI{6}{\micro\meter}) the Coulombian interaction between the particle surfaces is negligible.

The stiffness of one trap at equilibrium can be measured by calculating the variance of the x-displacement of the bead $\sigma^{2}_{x}$ which, because of  energy equipartition, is equal to $\frac{k_{\mathrm{B}}T}{k}$ where $k_{\mathrm{B}}$ is the Boltzmann constant and $T$ the temperature. Equivalently the power spectrum of the x-displacement is Lorentzian since the particles are over-damped: $S(f) = \frac{2\gamma k_{\mathrm{B}}T/k^2}{1+f^2/f_c^2}$, and one can fit it to find the cut-off frequency $f_c$ that verifies $f_c = \frac{k}{2\pi\gamma}$ where $\gamma = 6\pi R \eta$ and $\eta$ is the dynamic viscosity of water. The two methods give compatible results (assuming the viscosity of water and corrections due to the finite distance between the particle and the bottom of the cell are known).

To create an effective temperature on one of the particles (for example on particle $1$), a Gaussian white noise is sent to the AOD so that the position of the corresponding trap is moved randomly in the direction where the particles are aligned. If the amplitude of the displacement is sufficiently small to stay in the linear regime it creates a random force on the particles which does not affect the stiffness of the trap. Here the particles are over-damped and have a Lorentzian power spectrum with a typical cut-off frequency $f_c$ of \SI{30}{\hertz}. The added noise is numerically created by a LABVIEW program: it is sampled at \SI{100}{\kilo\hertz} with a tunable amplitude $A$ (typically of $\sim \SI{1}{\volt}$) and numerically low-pass filtered at \SI{1}{\kilo\hertz}. It is then generated by the analog output of a NI PXIe-6366 card. The conversion factor for the displacement due to the AOD is \SI{2.8}{\micro\meter/\volt}, and the typical voltage of the noise after filtration is between $\pm \SI{0.25}{\volt}$. When the random force is switched on, the bead quickly reaches a stationary state with an ``effective temperature'' for the randomly forced degree of freedom.

\begin{figure}[ht!]
\begin{center}
\includegraphics[width=8cm]{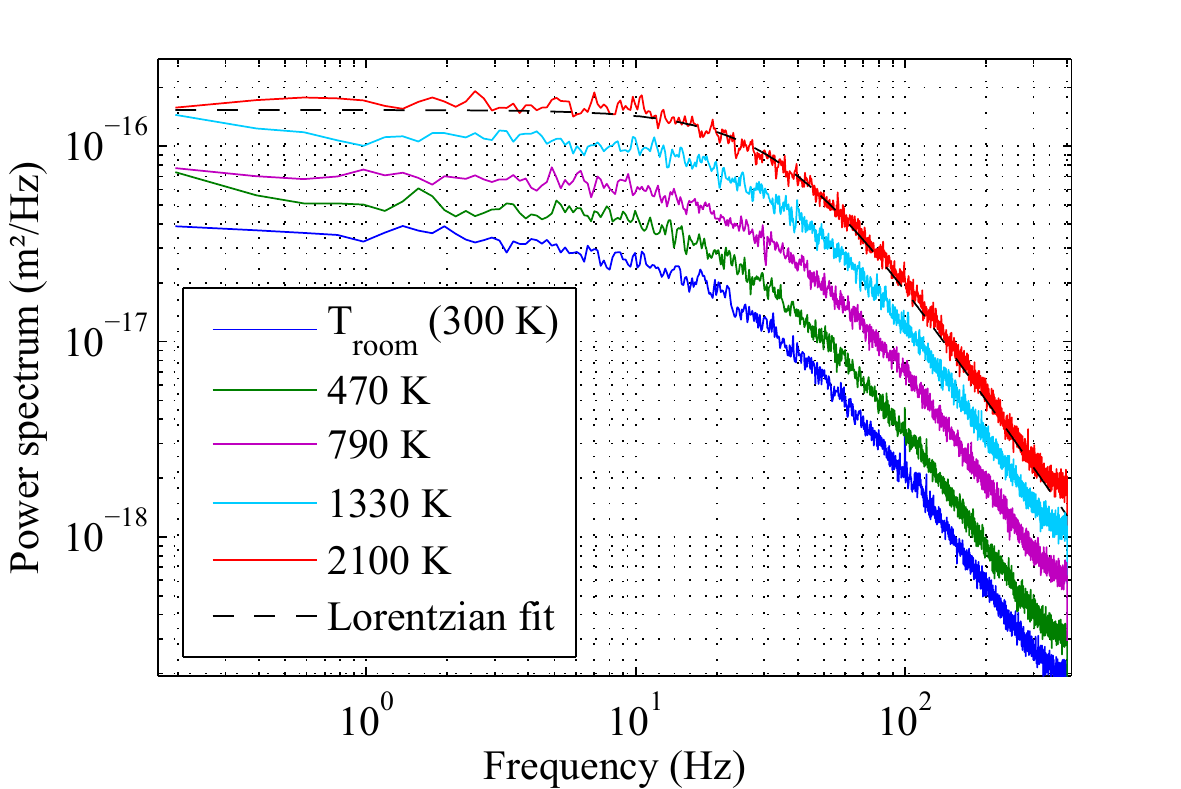}
\caption{Power spectra of the x-displacement of one bead of radius $R = \SI{1}{\micro\meter}$ trapped with stiffness $k = \SI{3.4}{\pico\newton/\micro\meter}$ in water at room temperature, at equilibrium (lowest blue curve), and for noise amplitude $A$ from $0.6$ to \SI{1.8}{\volt} ($A$ is incremented of \SI{0.4}{\volt} between each curve). The black dashed line is a Lorentzian fit of the spectrum with $A = \SI{1.8}{\volt}$. The indicated effective temperatures are calculated from the change of variance because  the stiffness and viscosity are not modified by the forcing.} \label{fig:powerspectrum}
\end{center}
\end{figure}

The power spectra of one bead's displacement in the x-direction with different noise amplitude (between $0$ and \SI{1.8}{\volt}) are shown in fig.\ref{fig:powerspectrum}. The displacement in the y-direction is not modified by the added noise.

As in \cite{ref:Petrov}, the power spectra when the bead is randomly forced are just vertical translations of the equilibrium one, which shows that only the effective temperature is modified (and not the stiffness of the trap, nor the viscosity of water). The cut-off frequency obtained by fitting the power spectra is not modified by more than a few hertz when the amplitude of the forcing is lower than \SI{1.5}{\volt}. For higher forcing amplitude, $f_c$ starts to be modified and the spectrum starts to be slightly less accurate at high frequency. This happens because the forced random displacement of the trap is too big compared to the size of the harmonic interval of the trapping potential.

This setup allows us to create a wide range of effective temperatures for one bead, and to look at the interaction between this agitated bead and another one trapped at equilibrium at a finite distance $d$.

\begin{figure}[ht!]
\begin{center}
\subfigure[]{
\includegraphics[width=8cm]{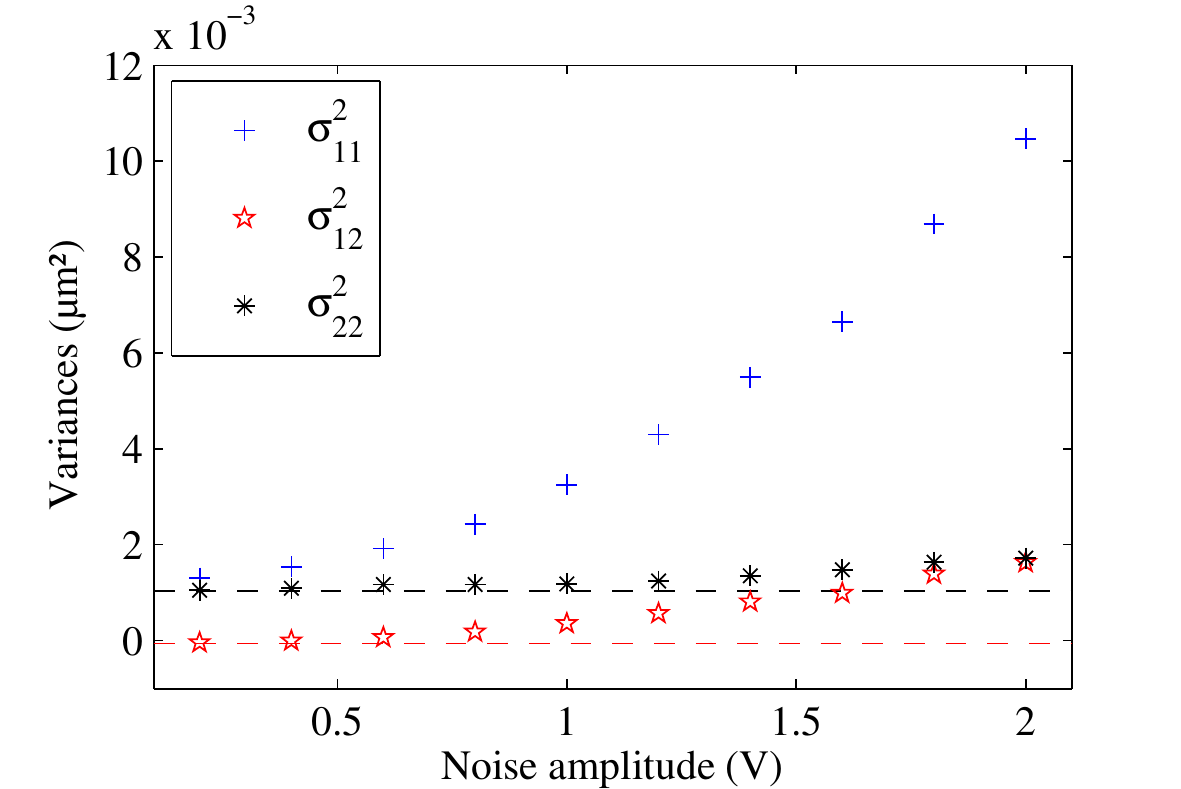}
\label{fig:variances_noise}}\\
\subfigure[]{
\includegraphics[width=8cm]{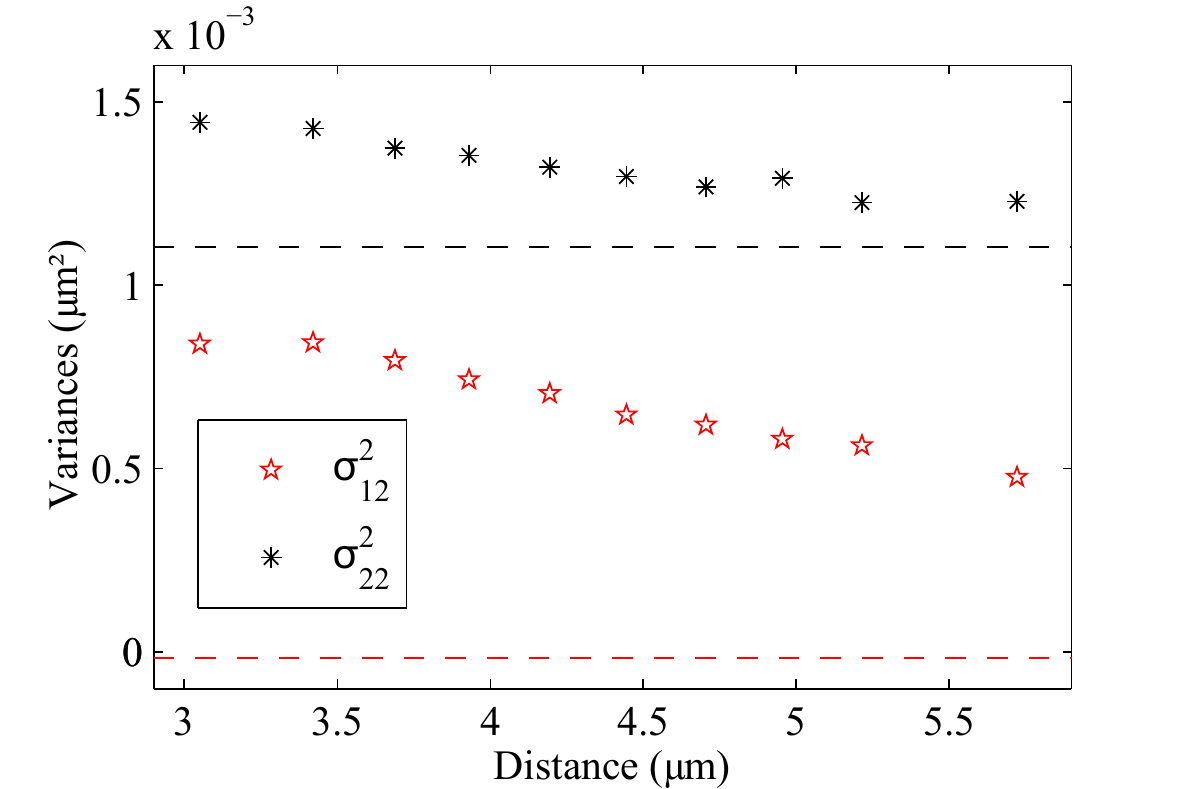}
\label{fig:variances_dist}}
\caption{Variance of the displacement of each bead ($\sigma^{2}_{22}$ and $\sigma^{2}_{11}$) and cross-variance between the two displacement ($\sigma^{2}_{12}$). (a) When the random forcing amplitude $A$ is increased on the first bead and the distance between the traps is kept constant, $d = \SI{3.2}{\micro\meter}$, the variances and the cross-variance increase. The dashed-lines are the values of $\sigma^{2}_{22}$ and $\sigma^{2}_{12}$ measured when there is no random forcing. (b) Zoom on $\sigma^{2}_{22}$ and $\sigma^{2}_{12}$ for a fixed forcing amplitude $A = \SI{1.5}{\volt}$, both values decrease with $d$ the mean distance between the two particles ($\sigma_{11}^2$ which is not shown remains nearly constant and equal to \SI{5.7d-3}{\square\micro\meter}). The dashed-lines are the values of $\sigma^{2}_{22}$ and $\sigma^{2}_{12}$ averaged over $d$ when there is no random forcing.} \label{fig:variances}
\end{center}
\end{figure}

When the first bead is forced we observe that the variance of its x-displacement $\sigma^{2}_{11} = \langle x_1 x_1 \rangle$ increases, which corresponds to the effect of the random forcing. The variance of the second particle's displacement $\sigma^{2}_{22} = \langle x_2 x_2 \rangle$ is also increased due to the coupling between the two particle, and more surprisingly the cross-variance $\sigma^{2}_{12} = \langle x_1 x_2 \rangle$ (which is the instantaneous cross-correlation of the x-displacements) ceases to be zero and increases with the amplitude of the random noise (see fig. \ref{fig:variances_noise}). For a fixed noise amplitude, the values of $\sigma^{2}_{22}$ and $\sigma^{2}_{12}$ also slightly decrease with the distance $d$ between the particles (see fig. \ref{fig:variances_dist}).

To understand this behaviour, we can use the classical hydrodynamic coupling. Following \cite{ref:Quake,ref:Bartlett,ref:Ou-Yang} the motion of two identical particles of radius $R$ trapped at positions separated by a distance $d$ is described by two coupled Langevin equations:
\begin{equation}
\left( \begin{array}{c} \dot{x}_1 \\ \dot{x}_2 \end{array} \right) = 
\mathcal{H}
\times \left( \begin{array}{c} F_1 \\ F_2 \end{array} \right)
\end{equation}
where $\mathcal{H}$ is the hydrodynamic coupling tensor, $x_i$ is the position of the particle $i$ relative to its trapping position and $F_i$ is the force acting on the particle $i$.

\noindent In the case where the displacements are small compared to the mean distance between the particles, the hydrodynamic coupling tensor reads:
\begin{equation}
\mathcal{H} = 
\begin{pmatrix} 1/\gamma & \epsilon/\gamma \\
 \epsilon/\gamma & 1/\gamma \end{pmatrix}
\end{equation}
where $\gamma$ is the Stokes friction coefficient ($\gamma = 6 \pi R \eta$ where $\eta$ is the viscosity of water) and $\epsilon$ is the coupling coefficient ($\epsilon = \frac{3R}{2d}$ if one takes the first order of the Oseen tensor \cite{ref:Doi}, $\epsilon = \frac{3R}{2d} - \left(\frac{R}{d}\right)^3$ if one takes the Rotne-Prager diffusion tensor \cite{ref:Herrera}).

\noindent At equilibrium the forces acting on the particles are:
\begin{equation}
F_i = -k_i \times x_i + f_i
\end{equation}
where $k_i$ is the stiffness of the trap $i$ and $f_i$ are the Brownian random forces which verify:
\begin{equation}
	\begin{array}{c}
		\langle f_i(t) \rangle = 0 \\
		\langle f_i(t) f_j(t^{\prime}) \rangle = 2k_{\mathrm{B}}T \, (\mathcal{H}^{-1})_{ij} \, \delta(t-t^{\prime})
	\end{array}
\end{equation}
where $k_{\mathrm{B}}$ is the Boltzmann constant and $T$ the temperature of the surrounding fluid.

Here we simply add an external random force $f^*$ on the first particle. This force is completely decorrelated with the Brownian random forces and characterised by an additional effective temperature $\Delta T$ (the particle $1$ is then at an effective temperature $T^* = T + \Delta T$).
\begin{equation}
	\begin{array}{c}
		\langle f^*(t) \rangle = 0 ~~\mbox{and}~~ \langle f^*(t) f_i(t^{\prime}) \rangle = 0 \\
		\langle f^*(t)f^*(t^{\prime}) \rangle = 2k_{\mathrm{B}} \Delta T \gamma \delta(t-t^{\prime})
	\end{array}
\end{equation}
\noindent It follows that the system of equations is:
\begin{equation}
\left\{
  \begin{array}{l}
    \gamma \dot{x}_1 = -k_1 x_1 + \epsilon (-k_2 x_2 + f_2) + f_1 + f^* \\
    \gamma \dot{x}_2 = -k_2 x_2 + \epsilon (-k_1 x_1 + f_1 + f^*) + f_2
  \end{array}
\right.
\label{eq:coupledLangevin}
\end{equation}

\noindent It can be rewritten:
\begin{equation}
\left\{
  \begin{array}{l}
    \dot{x}_1 = g_1(x_1,x_2) + \xi_1 \\
    \dot{x}_2 = g_2(x_1,x_2) + \xi_2
  \end{array}
\right.
\label{eq:coupledLangevinsimplified}
\end{equation}
with:
\begin{equation}
  \begin{array}{rcl}
    g_i(x_i,x_j) & = & -\frac{1}{\gamma} k_i x_i - \frac{\epsilon}{\gamma} k_j x_j  \\
    \xi_1 & = & \frac{1}{\gamma} (f_1 + \epsilon f_2 + f^*)\\
    \xi_2 & = & \frac{1}{\gamma} (f_2 + \epsilon f_1 + \epsilon f^*)
  \end{array}
\end{equation}

The equations are close to those describing the energy exchanged between two heat baths coupled by thermal fluctuations \cite{ref:Ciliberto} and it can be proved that the time evolution of the joint probability distribution function (PDF) $P(x_1,x_2,t)$ is governed by the Fokker-Planck equation \cite{ref:Zwanzig}:
\begin{eqnarray}
	\frac{\partial P}{\partial t} = & - \frac{\partial(g_1 P)}{\partial x_1} - \frac{\partial(g_2 P)}{\partial x_2} + 2 \theta _{12} \frac{\partial ^2 P}{\partial x_1 \partial x_2} \nonumber \\
	 & + \theta_{11} \frac{\partial ^2 P}{\partial x_1^2} + \theta_{22} \frac{\partial ^2 P }{\partial x_2^2}
\label{eq:FK}
\end{eqnarray}
where $\theta_{ij}$ is defined by:
\begin{equation}
\langle \xi_i(t) \xi_j(t^{\prime}) \rangle = 2 \theta_{ij} \delta(t-t^{\prime})
\end{equation}
\noindent Here we have:
\begin{equation}
  \begin{array}{l}
    \theta_{11} = k_{\mathrm{B}} (T+\Delta T) / \gamma \\
    \theta_{12} = k_{\mathrm{B}} \epsilon (T+\Delta T) / \gamma \\
    \theta_{22} = k_{\mathrm{B}} (T+\epsilon^2\Delta T) / \gamma
  \end{array}
\end{equation}

\noindent The stationary solution of equation \ref{eq:FK} can be written:
\begin{equation}
P_s(x1,x2)= \frac{\sqrt{ac-b^2}}{\pi} e^{-(ax_1^2+2bx_1x_2+cx_2^2)}
\end{equation}
where 
\begin{equation}
	\begin{array}{l}
		a = \frac{k_1 (k_1 + k_2) \left((k_1 + k_2) T + \epsilon^2 k_2 \Delta T\right)}{2 k_{\mathrm{B}} \left((T^2 + T \Delta T) (k_1 + k_2)^2 - \epsilon^2 (\epsilon^2-1) k_2^2 \Delta T^{2})\right)} \\
		b = \frac{- \epsilon k_1 k_2 (k_1 + k_2) \Delta T}{2 k_{\mathrm{B}} \left((T^2 + T \Delta T) (k_1 + k_2)^2 - \epsilon^2 (\epsilon^2-1) k_2^2 \Delta T^{2})\right)}\\
		c = \frac{k_2 (k_1 + k_2) \left((k_1 + k_2) T + (k_1 + k_2 (1- \epsilon^2)) \Delta T)\right)}{2 k_{\mathrm{B}} \left((T^2 + T \Delta T) (k_1 + k_2)^2 - \epsilon^2 (\epsilon^2-1) k_2^2 \Delta T^{2})\right)}
	\end{array}
\end{equation}

\begin{figure}[ht!]
\begin{center}
\subfigure[]{
\includegraphics[width=8cm]{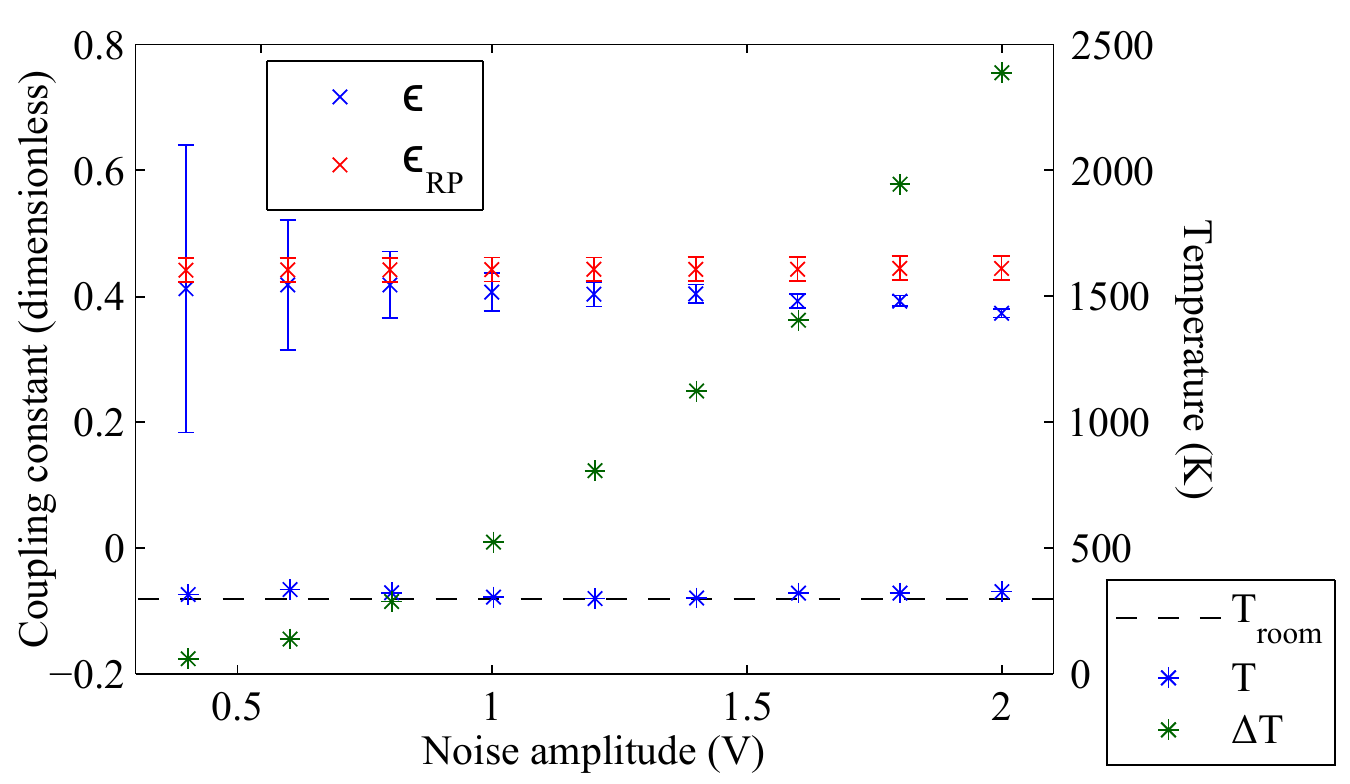}
\label{fig:diff_amp}}\\
\subfigure[]{
\includegraphics[width=8cm]{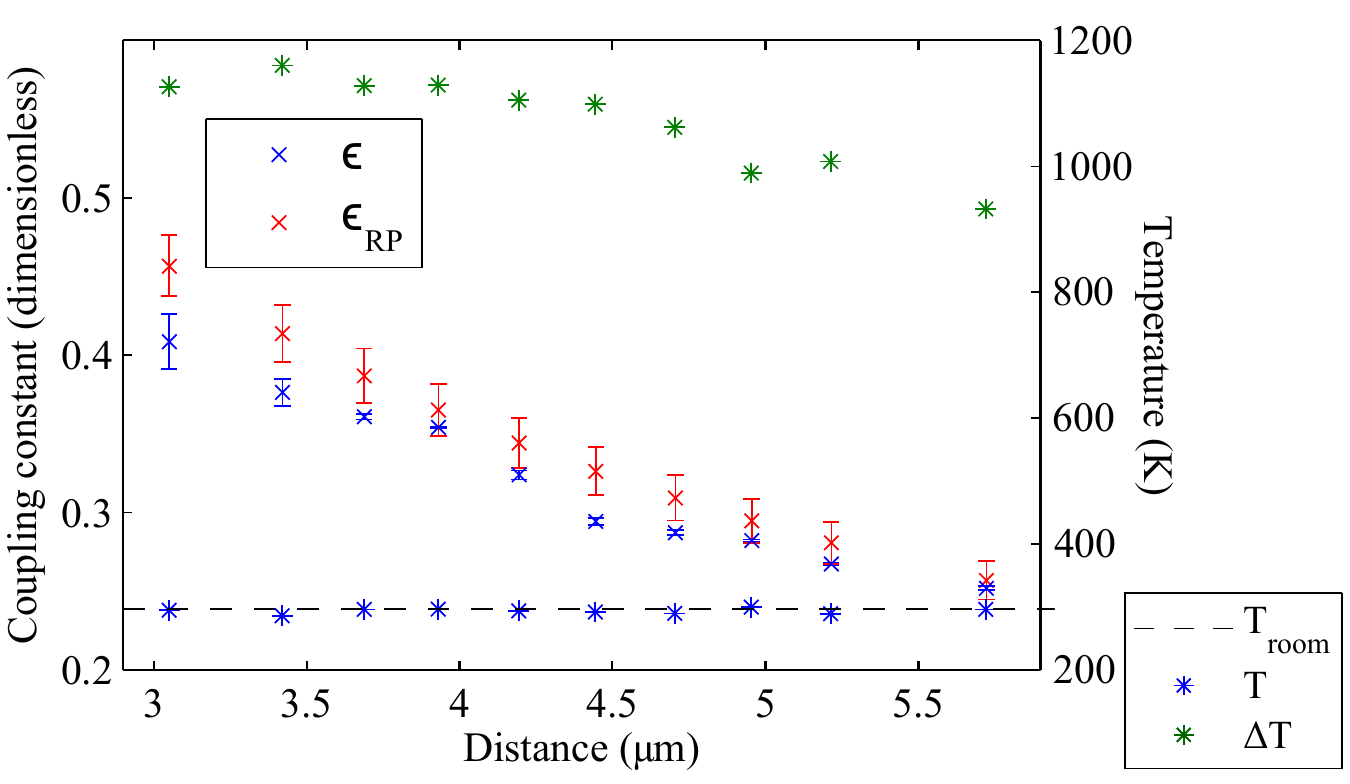}
\label{fig:diff_dist}}
\caption{Coupling coefficient ($\epsilon$), temperature of the bath ($T$) and effective temperature ($\Delta T$), measured from the values of $\sigma^{2}_{11}$,$\sigma^{2}_{12}$ and $\sigma^{2}_{22}$, and theoretical coupling coefficient from the Rotne-Prager diffusion tensor ($\epsilon_{\mathrm{RP}}$) for particles of radius $R = \SI{1}{\micro\meter} \pm 5\%$. (a) For two particles trapped at distance $d=\SI{3.2}{\micro\meter}$ as a function of the amplitude $A$ of the forcing done on one particle. (b) For two particles at different effective temperature as a function of the distance $d$ between the particles.}
\end{center}
\end{figure}

Then, one can compute the variances of each position and the cross-variance between the two particles:
\begin{equation}
\begin{array}{l}
\sigma^{2}_{11} = \langle x_1 x_1 \rangle = \frac{k_{\mathrm{B}}(T+\Delta T)}{k_1} - \frac{k_2}{k_1}\frac{\epsilon ^2 k_{\mathrm{B}} \Delta T}{k_1+k_2} \\
\sigma^{2}_{12} = \langle x_1 x_2 \rangle = \frac{\epsilon k_{\mathrm{B}} \Delta T}{k_1 + k_2} \\
\sigma^{2}_{22} = \langle x_2 x_2 \rangle = \frac{k_{\mathrm{B}}T}{k_2} + \frac{\epsilon ^2 k_{\mathrm{B}} \Delta T}{k_1+k_2}
\end{array}
\label{eq:variances}
\end{equation}
\noindent This result shows the appearance of the non-zero cross-variance which does not exist in the equilibrium case, and an exchange of energy between the two particles. Indeed the variances can be rewritten $\sigma^{2}_{11} = \sigma^{2}_{1 \, \mathrm{n.c.}} - \frac{k_2}{k_1}\frac{\epsilon ^2 k_{\mathrm{B}} \Delta T}{k_1+k_2}$ and $\sigma^{2}_{22} = \sigma^{2}_{2 \, \mathrm{n.c.}} + \frac{\epsilon ^2 k_{\mathrm{B}} \Delta T}{k_1+k_2}$ where $\sigma^{2}_{i \, \mathrm{n.c.}}$ is the variance of the particle $i$ with no  coupling. It follows that the variance of the ``hot'' particle (the forced one) is decreased by the presence of the ``cold'' particle, and reciprocally the variance of the cold one is increased by the presence of the hot one.

By measuring $\sigma^{2}_{11}$,$\sigma^{2}_{12}$ and $\sigma^{2}_{22}$, one can solve the system \ref{eq:variances} and find the values of $T$, $\Delta T$ and $\epsilon$ (given that $k_1$ and $k_2$ are measured separately). Some experimental values for a given distance $d$ and different amplitudes of forcing $A$ done on the particle $1$ are shown in figure \ref{fig:diff_amp}, and for a given forcing amplitude  and different distances are shown in figure \ref{fig:diff_dist}. As expected, $T$ is always nearly constant and equal to room temperature (all values are compatible with room temperature of $\SI{297}{\kelvin}$ with a precision of $10 \%$), $\epsilon$ depends only on the distance between the particles (in figure \ref{fig:diff_amp} all values are between $0.37$ and $0.42$), and $\Delta T$ depends only on the forcing amplitude done on the first particle.

In fig. \ref{fig:diff_amp} and \ref{fig:diff_dist} we notice that the measured value of $\epsilon$ is always slightly lower than the theoretical one (estimated by the Rotne-Prager diffusion tensor) but shows the same dependence in the distance $d$ between the two particles.\footnote{For this discrepancy it has been verified that the value of $\epsilon$ is not significantly modified if the distance between the bead and the bottom surface of the cell is changed to \SI{10}{\micro\meter} or \SI{20}{\micro\meter}.} Note that there are two experimental problems : a) for very low forcing (i.e. low $\Delta T$), the errorbars on $\epsilon$ are big because they are estimated considering that the main source of incertitude is the value of $\sigma^{2}_{12}$, which is very low when forcing is low.\footnote{Here the values of $\sigma^{2}_{12}$ used for computation are corrected by subtracting the value of the cross-variance when the system is at equilibrium (this value should theoretically be zero and gives an estimation of the incertitude on $\sigma^{2}_{12}$).} b) when the forcing is very high, the estimation of $\epsilon$ starts to be less precise because, as already mentioned, the added random force begins to be less accurate for high displacements of the trap position. In fig. \ref{fig:diff_dist} also the effective temperature $\Delta T$ slightly decreases when the distance $d$ is increased because of the less accurate response of the AOD far from the center of the apparatus\footnote{The shape of the trap is always impaired when the beam is not well centred, which lowers the stiffness of the trap and the $\Delta T$ corresponding to a given noise amplitude.}.

\begin{figure}[ht!]
\begin{center}
\includegraphics[width=8cm]{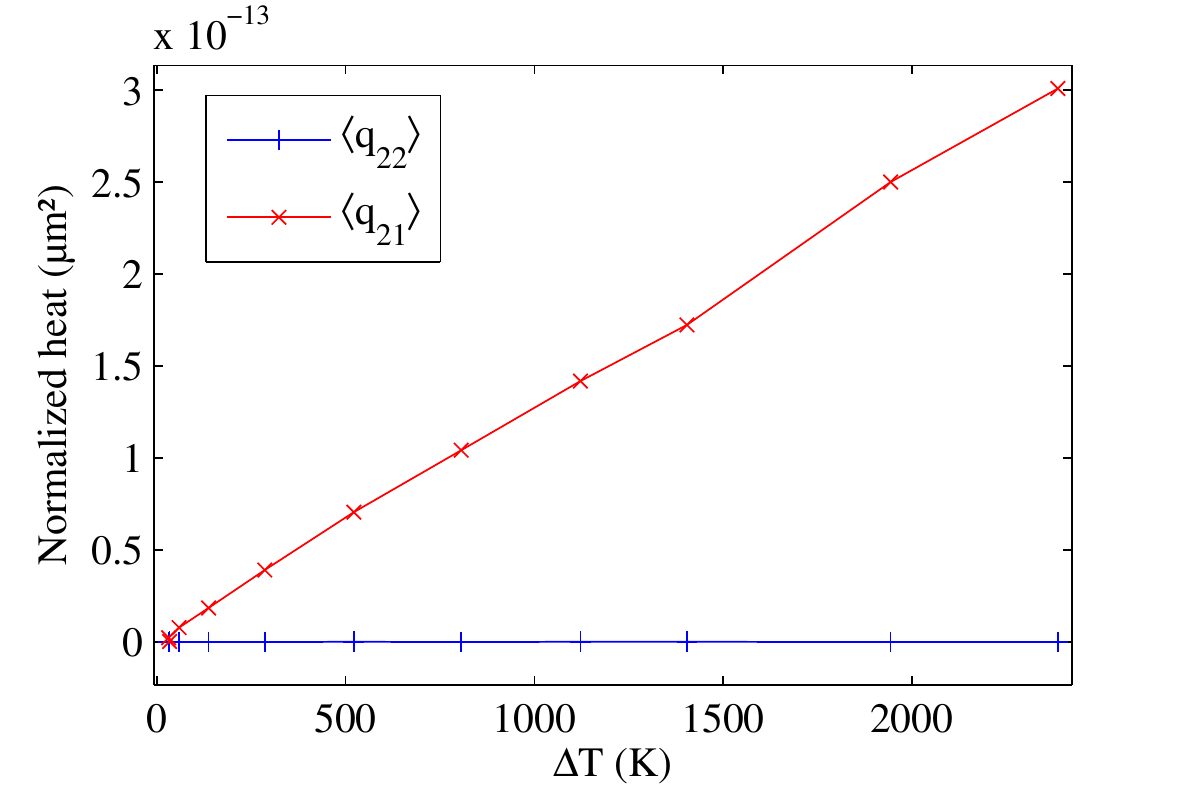}
\caption{Mean normalized heat received by particle $2$ during a time $\tau = \SI{1}{\second}$ ($d = \SI{3.2}{\micro\meter}$). The average is done on \SI{500}{} independent portions of trajectories. {The mean normalized heat dissipated by particle $1$ is not shown because the curves are too close to be differentiated.}}
\label{fig:heats}
\end{center}
\end{figure}

It is interesting to notice that the values of $\sigma^{2}_{11}$, $\sigma^{2}_{12}$ and $\sigma^{2}_{22}$ are linked to the mean heat flux between the two particles. Indeed the heat dissipated\footnote{In the published version we have written ``received'' but it was a mistake.} by the particle $i$ during the time $\tau$ is given by \cite{ref:Sekimoto}:
\begin{equation}
Q_i (\tau) = \int_{0}^{\tau} \left( \gamma \dot{x}_i - \gamma \xi_i \right) \dot{x}_i \, \mathrm{d}t
\end{equation}
Using equations \ref{eq:coupledLangevinsimplified} it can be decomposed in two terms:
\begin{equation}
Q_i (\tau) = k_i q_{ii} + \epsilon k_j q_{ij}
\end{equation}
Where:
\begin{equation}
\begin{array}{l}
q_{ii} = - \int_{0}^{\tau} x_i \dot{x}_i \, \mathrm{d}t \\
q_{ij} = - \int_{0}^{\tau} x_j \dot{x}_i \, \mathrm{d}t 
\end{array}
\end{equation}
The average of the two terms $q_{22}$ and $q_{21}$ contributing to $Q_2$, integrated over \SI{1}{\second}, are shown figure \ref{fig:heats} for different effective temperatures $\Delta T$. These values are very close to the opposite of the terms contributing to $Q_1$. The maximal difference between $-\langle q_{12} \rangle$ and $\langle q_{21} \rangle$ is of \SI{0.24}{\percent} and both terms depends linearly on $\Delta T$. Moreover, the terms $\langle q_{11} \rangle$ and $\langle q_{22} \rangle$ are always nearly equal to zero (the maximal value observed is \SI{6d-16}{\square\micro\meter}), which is normal since $\int_{0}^{\tau} - x_i \dot{x}_i \, \mathrm{d}t = - \left[ \frac{1}{2} x_i^2 \right]_0^{\tau}$. Then the mean heat dissipated by particle $i$ during time $\tau$ is:
\begin{equation}
\langle Q_i (\tau) \rangle = \epsilon k_j \langle q_{ij} \rangle
\end{equation}
It follows that the mean dissipated heat by particle $1$ and received heat by particle $2$ are proportional to $\Delta T$ as would be a normal mean heat flux between two sources at different temperatures. This result allows us to interpret the cross-variance $\sigma^{2}_{12}$ and the difference $\sigma^{2}_{ii} - \sigma^{2}_{i \, \mathrm{n.c.}}$, which also depend linearly on $\Delta T$, as proportional to the heat flux going from the particle $1$ (``hot'') to the particle $2$ (``cold'').

Finally, following the resolution method described in \cite{ref:Ou-Yang} and using the eqs. \ref{eq:variances}, one can compute the cross-correlation functions of $x_1$ and $x_2$ for time $t$ ($t>0$):
\begin{multline}
{\textstyle \langle x_1(t) x_2(0) \rangle = \frac{\epsilon k_{\mathrm{B}}}{2(k_1+k_2) \kappa} \times }\\
{\scriptstyle  \left[ \left(\Delta T(\kappa+k_1+k_2(2\epsilon^2-1))+2T(k_1+k_2)\right) e^{- \frac{((k_1+k_2)-\kappa)t}{2\gamma}} \right. }\\
{\scriptstyle \left. + \left(\Delta T(\kappa-k_1-k_2(2\epsilon^2-1))-2T(k_1+k_2)\right) e^{- \frac{((k_1+k_2)+\kappa)t}{2\gamma}} \right] }
\label{eq:x1x20_dif}
\end{multline}
\begin{multline}
{\textstyle \langle x_1(0) x_2(t) \rangle = \frac{\epsilon k_{\mathrm{B}}}{2(k_1+k_2) \kappa} \times }\\
{\scriptstyle \left[ \left(\Delta T(\kappa+k_1+k_2(3-2\epsilon^2))+2T(k_1+k_2)\right) e^{- \frac{((k_1+k_2)-\kappa)t}{2\gamma}} \right. }\\ 
{\scriptstyle \left. + \left(\Delta T(\kappa-k_1-k_2(3-2\epsilon^2))-2T(k_1+k_2)\right) e^{- \frac{((k_1+k_2)+\kappa)t}{2\gamma}} \right] }
\label{eq:x2x10_dif}
\end{multline}
with :
\begin{equation}
\kappa = \sqrt{k_1^2-2k_1k_2+k_2^2+4\epsilon^2k_1k_2}
\end{equation}
\noindent When $k_1=k_2=k$ the expressions can be simplified:
\begin{multline}
{\textstyle \langle x_1(t) x_2(0) \rangle = \frac{k_{\mathrm{B}}}{4k} \times }\\
{\scriptstyle \left[ \left(-2T+\Delta T \epsilon(1-\epsilon)\right) e^{-\frac{k(1-\epsilon)t}{\gamma}} + \left(2T+\Delta T \epsilon(1+\epsilon) \right) e^{-\frac{k(1+\epsilon)t}{\gamma}} \right]}
\label{eq:x1x20_same}
\end{multline}
\begin{multline}
{\textstyle \langle x_1(0) x_2(t) \rangle = \frac{k_{\mathrm{B}}}{4k} \times }\\
{\scriptstyle \left[ \left(-2T+\Delta T (-2 + \epsilon + \epsilon^2)\right) e^{-\frac{k(1-\epsilon)t}{\gamma}}  +  \right.} \\
{\scriptstyle \left. \left(2T+\Delta T (2+\epsilon-\epsilon^2)\right) e^{-\frac{k(1+\epsilon)t}{\gamma}} \right]}
\label{eq:x2x10_same}
\end{multline}
\noindent {Of course if $\Delta T = 0$, $\langle x_1(0) x_2(t) \rangle$ and $\langle x_1(0) x_2(t) \rangle$ are equal because the two beads play the same role and the expressions become the same as the ones obtained in \cite{ref:Quake,ref:Bartlett,ref:Ou-Yang}.}
\begin{figure}[ht!]
\begin{center}
\subfigure[]{
\includegraphics[width=8cm]{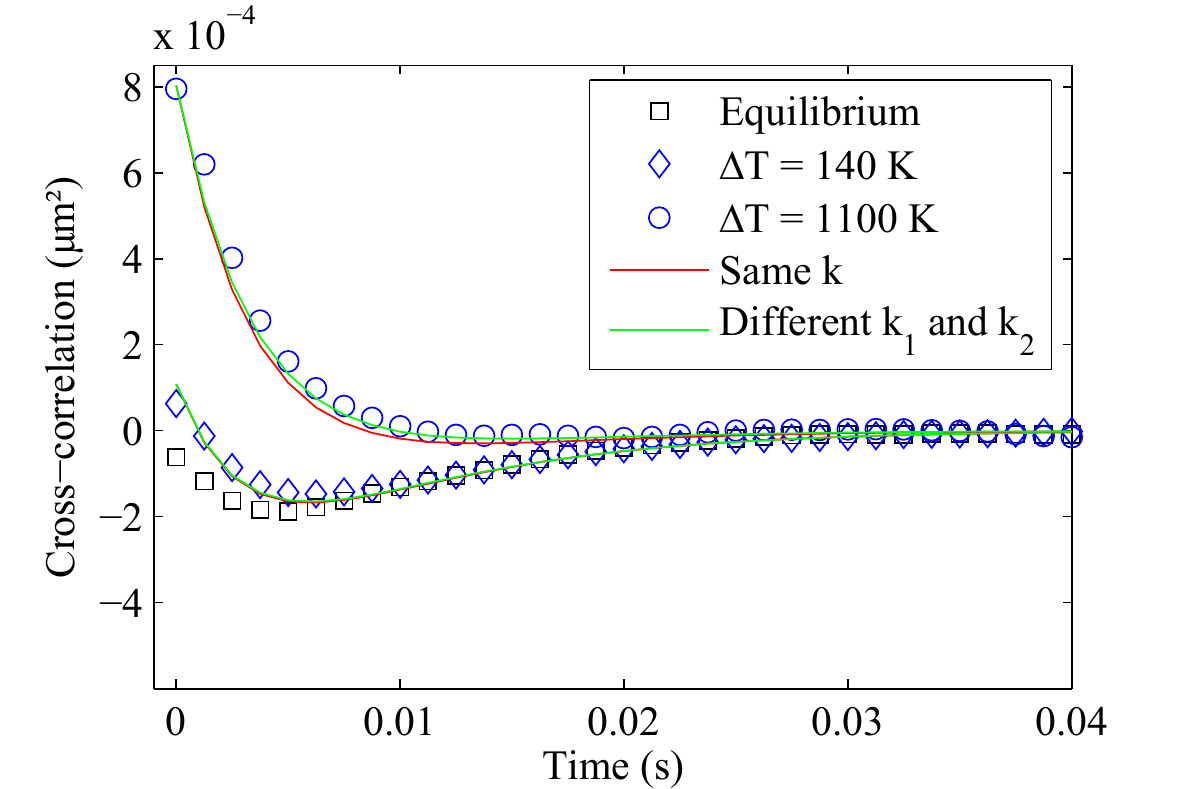}}\\
\subfigure[]{
\includegraphics[width=8cm]{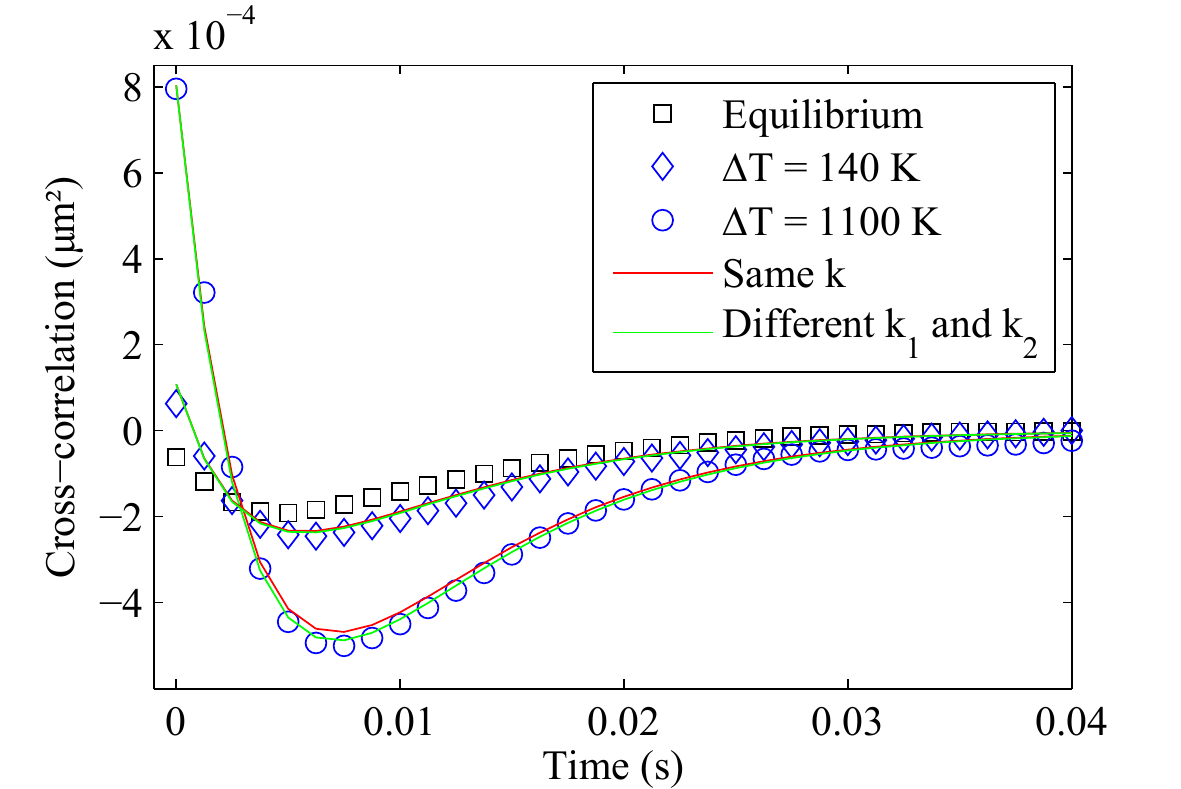}}
\caption{Measured cross-correlations compared to the theoretical expressions with measured parameters, by taking account of the slightly different values of $k_1$ and $k_2$ (green) or considering a unique mean value for the two stiffness $k_1 = k_2 = k$ (red). (a) $\langle x_1(t) x_2(0) \rangle$. (b) $\langle x_1(0) x_2(t) \rangle$.}
\label{fig:crosscorr}
\end{center}
\end{figure}
The theoretical expressions of the cross-correlation functions can be compared with the experimental data since all parameters can be measured. The results are shown figure \ref{fig:crosscorr}. The data show a good agreement with the model.  Since the values of $k_1$ and $k_2$ are nearly equal (for the data shown figure \ref{fig:crosscorr}: $k_1 \simeq \SI{3.4}{\pico\newton/\micro\meter}$ and $k_2 \simeq \SI{4.0}{\pico\newton/\micro\meter}$) there is no big difference between the curves obtained from eqs. \ref{eq:x1x20_dif}-\ref{eq:x2x10_dif} (green curves) and those  obtained from eqs. \ref{eq:x1x20_same}-\ref{eq:x2x10_same} using for k the mean value of $k_1$ and $k_2$ (red curves). Note that, contrary to the equilibrium case, $\langle x_1(0) x_2(t) \rangle$ and $\langle x_1(t) x_2(0) \rangle$ are not equal, since the roles of particles $1$ and $2$ are not symmetrical. $\langle x_1(0) x_2(t) \rangle$ always shows a time-delayed anti-correlations more pronounced than in the equilibrium case whereas $\langle x_1(t) x_2(0) \rangle$ doesn't show any anti-correlation as soon as $\Delta T \geq \frac{2}{\epsilon (1 - \epsilon)}T$. This behaviour can be understood in the following way: $ \langle x_i(0)x_j(t) \rangle $ is linked to the influence that $x_i$ at a given time $t=0$ has on $x_j$ after a time $t$. Since $x_1$ is forced, it is less sensitive to the motion of $x_2$, whereas $x_2$ is more sensitive to the motion of $x_1$ which is bigger than its own motion.\bigskip


In conclusion, we have shown that the random forcing of the position of trapped bead does not modifies the trap stiffness and it can be interpreted  as an effective temperature for the bead. This effective temperature has been used to study the energy fluxes and the correlations functions  between two particles at different temperatures and coupled only by hydrodynamic interactions.  The main result of this letter is that these interactions,  simply described by the classical hydrodynamic coupling tensor, gives rise to an unusual instantaneous cross-correlation between the motions of the particles and an effective energy exchange from the hot bead to the cold bead, which are proportional to the mean heat flux between the two particles. The experimental results are in very good agreement with the prediction of a theoretical model based on the resolution of two coupled Langevin equations, using equivalent Fokker-Planck equations.

%


\end{document}